# The role of temporal cortex in the control of attention

Hamidreza Ramezanpour[1,2,3] and Mazyar Fallah[1,2,3,4,5]

[1] Centre for Vision Research, York University, Toronto, Ontario, Canada
[2] School of Kinesiology and Health Science, Faculty of Health, York University, Toronto, Ontario, Canada
[3] VISTA: Vision Science to Application, York University, Toronto, Ontario, Canada
[4] Department of Psychology, Faculty of Health, York University, Toronto, Ontario, Canada
[5] Department of Human Health and Nutritional Sciences, College of Biological Science, University of Guelph, Guelph, Ontario, Canada

Correspondence should be addressed to: Hamidreza Ramezanpour (hamidram@yorku.ca) or Mazyar Fallah (mfallah@uoguelph.ca).

## Abstract

Attention is an indispensable component of active vision. Contrary to the widely accepted notion that temporal cortex processing primarily focusses on passive object recognition, a series of very recent studies emphasize the role of temporal cortex structures, specifically the superior temporal sulcus (STS) and inferotemporal (IT) cortex, in guiding attention and implementing cognitive programs relevant for behavioral tasks. The goal of this theoretical paper is to advance the hypothesis that the temporal cortex attention network (TAN) entails necessary components to actively participate in attentional control in a flexible task-dependent manner. First, we will briefly discuss the general architecture of the temporal cortex with a focus on the STS and IT cortex of monkeys and their modulation with attention. Then we will review evidence from behavioral and neurophysiological studies that support their guidance of attention in the presence of cognitive control signals. Next, we propose a mechanistic framework for executive control of attention in the temporal cortex. Finally, we summarize the role of temporal cortex in implementing cognitive programs and discuss how they contribute to the dynamic nature of visual attention to ensure flexible behavior.

## Highlights

- The temporal cortex is not just a passive analyzer of sensory information to facilitate object recognition.
- When cognitive control signals interact with attentional control, a temporal cortex attention network (TAN) is  engaged to support flexible behavior.
- Active vision by temporal cortex generalizes the existence of attention controllers and priority maps to non-oculomotor structures.

- Functional properties of the TAN can be beneficial for action understanding and social interactions.

## Keywords

## Visual Attention

The brain can only process a limited amount of information received by our sensory system at a given point in time. Attention selects behaviorally relevant information to overcome this bottleneck. Numerous studies have investigated neural and behavioral effects of visual attention in humans and nonhuman primates [1–8]. In this paper, we primarily focus on the functional architecture of temporal cortex in nonhuman primates, where it is better understood, with comparison to similar findings in human studies.

Attention can be allocated either toward a specific spatial location (spatial attention), toward non-spatial features (feature-based attention) such as motion direction, color, or towards an object as defined by a combination of features at a location in the visual scene (object-based attention). Each of these types of attention has been shown to influence encoding of task-relevant locations or features throughout the visual cortical hierarchy [9–14] as well as higher-order areas such as the lateral intraparietal (LIP) area [15–17], frontal eye field (FEF) [18], and prefrontal cortex (PFC) [19,20].

The premotor theory of attention postulated that the neural networks involved in eye movement control (the oculomotor system) and attentional control do not differ and visual attention is a consequence of action planning [21]. This theory has received support from electrophysiological studies on monkeys [22–28] as well as neuroimaging studies in humans [29,30]. The most substantial support comes from a series of work on the monkey FEF, a critical control area for the oculomotor system, which showed that subthreshold stimulation of a motor vector, while not evoking a saccade, deploys attention to the location in the visual field that represents the endpoint of the motor plan [26–28]. This was followed up by a similar study in the superior colliculus, a subcortical oculomotor area [23]. Notwithstanding these pieces of evidence, some other studies have dissociated the coupling between endogenous attention and eye movements [31,32]. The limitation of premotor theory is that it is action-based and primarily driven by external stimuli.

In addition to the premotor theory of attention, several other theories such as biased competition [4,33], selective tuning [34,35], and the rhythmic theory of attention [36] have tried to encompass broader aspects of visual attention including both spatial and non-spatial (feature-based) attention. A detailed description of all of those models is beyond the scope of this review. Nevertheless, here we briefly review the concept of the selective tuning model [37] which is relevant for temporal cortex participation in attentional control. This model is based on the brain's hierarchical organization and assumes three stages: first, the stimuli enter the first layer and propagate to the upper layers via feed-forward connections in an inverted sub-pyramid manner. Second, a "winner take all" [38] process is applied to each layer of the network, starting from the output layer and back-propagating towards the input layer. Consequently, at each layer, the connections that are not contributing to the winner are pruned away. The pruned connections form a suppressive annular region around the connections that remain active and form the attention zone that selects parts of the stimuli [39,40]. Finally, in the third stage, the selected parts of the stimuli in the input layer propagate once again towards the output layer, but this time as if there are no distractors [41]. The selective tuning model is of interest here because it will enable us to explain the nature of temporal cortex participation in attentional control and how it might depend on the stimulus position in the visual field. The selective tuning model also fits with the findings of the recent studies on the temporal cortex [42–46] and supports the integration of cognitive programs which will be discussed in the next sections, but that does not automatically discount other models of visual attention.

Many of these models of selective attention are dependent upon saliency and feature conspicuity maps. It has been hypothesized every visual scene can be segmented into separate feature conspicuity maps [38,47] representing a single feature, such as orientation or color [38,48,49]. These feature conspicuity maps are topographically organized and compete for selection. Topographic representation of the weighted sums of feature conspicuity map activations may generate a single master map (or "central representation"[38]) representing saliency in a passive (bottom-up) manner [38,50]. The observer's goal does not have any role in this type of processing. The salience map is a theoretical framework, and its neural correlate (if it exists at all) remains an open question. Nevertheless, studies have tried to find the locus of the saliency map in the brain [51–62]. While the concept of saliency and conspicuity maps can help us understand many aspects of visual attention, behavior is best represented by a combination of the feature conspicuity maps with top-down relevance signals which reflect an active observer's goal. Hitherto, studies have suggested that these integrations mostly happen in oculomotor structures, i.e., FEF, SC, LIP (see ref [48] for review).

The various models of visual attention just mentioned are generally mutually inconsistent even though each gives a supportable perspective on the overall attention problem. In this paper, we review recent findings which suggest that temporal cortex regions act on feature conspicuity representations, contribute to guiding attention, and implement cognitive control signals with regard to one of these theories, i.e. the selective tuning model, already set up to encompass these elements [37,63,64] in a novel manner.

## Visual processing in the STS and IT cortex

The temporal cortex can be coarsely divided into four sub-regions: medial temporal cortex (MTC), superior temporal gyrus (STG), superior temporal sulcus (STS) and inferotemporal (IT) cortex (see [65] for a review). In this review, we are focusing on the STS and IT, as the MTC and STG have yet to evidence support for attentional control. Classical studies have considered two separate but parallel streams for processing of visual information: the dorsal stream (where pathway) dealing with spatial aspects of stimuli, and the ventral stream (what pathway) implicated in recognition of object [66,67]. Later theories emphasized that the dorsal stream information is used to guide actions while ventral stream processing is necessary for perception [68]. For this review, we are focusing on the ventral stream which includes areas V1, V2, V4, and IT cortex (TEO and TE in nonhuman primate studies). Processing of the visual input becomes progressively more complex along the ventral stream, taking form and color information to produce object recognition [69–72].

IT cortex is coarsely divided into three subregions: posterior IT (pIT), central IT (cIT) and anterior IT (aIT) [73,74]. Based on cytoarchitectural divisions, pIT corresponds to TEO while cIT and aIT correspond to TE [73]. Early neurophysiological studies investigated stimulus selectivity and receptive fields properties of IT cortex neurons [75,76], determining that IT neurons have large bilateral receptive fields and complex object selectivity, including faces. While single-unit recordings showed that there is a columnar organization for shape processing [77] in the IT cortex, later neurophysiological and neuroimaging studies led to the discovery of several patches within IT cortex selective for specific categories of objects such as faces, houses, or more complex stimuli including scenes [78–82]. Neuroimaging studies in humans also revealed other potential principles of the IT cortex, namely retinotopic organization and real-world size representation of objects [83,84]. However, until very recently, the general principle governing IT cortex organization was unclear. In a comprehensive study combining fMRI, electrophysiology, microstimulation, and deep networks, Bao and colleagues put forward a unified theory of IT cortex organization by showing that monkey IT cortex is topographically organized into a map of low dimensional object space that is repeated three times with increasing invariance at each stage [85].

In contrast, the superior temporal sulcus, which sits between the ventral and dorsal streams, is involved in a wide variety of functions [86,87], such as motion processing [88–90], speech processing [91], audiovisual integration [92,93], multisensory perception [94,95], and social interaction processing [96–98]. While the exact correspondence between various anatomical regions of the monkey and the human STS has not yet been established, most of the above-mentioned functions are shared between the two species. While it has been shown that complex information processing in the STS is partially handled by a number of specialized modules [86], this does not preclude the fact that the multifunctionality of the STS might be due to its coactivation with distinct neural networks implicated in distinct tasks [87]. The latter notion gets further support from the massive bidirectional connections of the STS with a range of higher-order brain areas such as the ventral and medial frontal cortex, lateral prefrontal and premotor areas, the parietal cortex, and mesial temporal regions [99,100] strategically placing it as a functional link between early visual areas and higher order areas. For example, integrating different modalities of information, such as vision and audition, is highly beneficial for disambiguating social decision processes [101]. Another example is facial expression recognition which builds upon the ability to combine biological motion and facial information [102]. In summary, the STS shows a wide functionality around integrating and linking complex information from early sensory cortices to be later processed by higher order areas.

## STS and IT cortex: recipients of visual attention

Although some studies have shown that the effects of attention are wide spread and impact most of the visual areas starting from the lateral geniculate nucleus (LGN) to the higher order temporal cortex areas such as TEO [103], the influence of attention on the functioning of the temporal cortex has been less elucidated than for early visual areas such as V2 and V4. Nevertheless, attention has been shown to modulate shape processing responses in the human STS [104]. For example, when dynamic aspects of human faces such as gaze or expressions are attended, neural responses of the pSTS face responsive areas are modulated [105–107]. One of the seminal studies of how attention influences the responses of ventral stream neurons was carried out by Moran and Desimone [108]. They recorded the response of neurons in areas V4 and the IT cortex of monkeys while presenting two visual stimuli simultaneously inside the neuron's receptive field. They found that the response to the unattended stimulus was dramatically reduced. Similar results were later reported in areas such as V2, V4, MT, MST and IT [6,109,110]. Visual attention was shown to increase neuronal responses without changing selectivity, which is described as gain modulation [14]. Around the same time, other studies showed that attention not only affects neural responses but also improves behavioral performance during visual discrimination or visual search tasks [108,111–113].

As visual areas were shown to be affected by visuospatial attention, further studies looked to determine where the attentional signals originated from. Studies have suggested that the prefrontal cortex is the primary driver of the temporal cortex during object recognition [114–116]. In the search for the source of attention control signals in the prefrontal cortex that drive object recognition specifically, two studies emphasized the role of inferior frontal junction (IFJ) in humans and its homolog in monkeys, ventral pre-arcuate gyrus (VPA). Using fMRI simultaneously with magnetoencephalography (MEG) in humans, Baldauf and Desimone tested the hypothesis that attention to different object categories may synchronize areas representing those categories in the temporal cortex and higher-order areas in the prefrontal cortex [117]. They used a sequence of stimuli consisting of two object categories superimposed (faces and houses), fading in and out of a phase–scrambled noise mask at different frequencies. Depending on which object was attended, an area close to Brodmann areas 45 and 46, the IFJ, was found to be synchronously activated with the fusiform face area (FFA) and parahippocampus place area (PPA) which represent faces and places respectively. The phase analysis showed that the synchronized gamma phases in the IFJ were advanced by 20 ms compared to FFA and PPA, suggesting that the IFJ is the synchrony driver in a top-down manner. In an experiment using electrophysiology and pharmacological manipulations in monkeys, Bichot and colleagues showed that VPA plays a similar role in visual search to control non-spatial attention [118]. Simultaneous recordings from VPA, FEF and IT cortex revealed that the VPA shows the earliest time of feature selection. Therefore, they concluded that the VPA must be the source of the feature selection. Supporting parallel mechanisms for the control of spatial and non-spatial attention, pharmacological deactivation of the VPA impaired feature selection but, contrary to the effects of FEF inactivation, had no significant effect on spatial attention [118]. Taken together, in contrast to the oculomotor network-driven spatial attention signals, the above-mentioned studies demonstrated that attentional modulations seen in IT cortex only contribute to object recognition. Nevertheless, despite all of these recent advancements in our understanding of the operations in IT cortex, most of the previous studies have neglected its potential role beyond object recognition through fast feedforward converging processing [69]. To be more specific, the possibility that IT cortex, or more generally speaking the temporal cortex structures, is an active participant in attentional control beyond simply being a target of such control has rarely been considered.

**Temporal cortex participates in attentional control**

One of the first studies showing that the IT cortex is involved in attentional selection was carried out by Rohtblat and Pribram [119]. They trained monkeys on a task which demanded selecting either the color or the form of a complex stimulus. They showed that initially the neural responses from IT cortex time-locked to the response, and as it was not stimulus-locked, could play a role in stimulus selection. However, the occipital cortex also showed selective responses to the attended dimension which were instead locked to the stimulus onset. These results shed further light on the mechanism by which attention can lead to selection: early input filtering via the occipital cortex as a consequence of learning or later response filtering via the IT cortex [119] (see **Figure 1**). It is this latter mechanism that suggests areas within IT would provide a basis for guiding attention and shaping behavior. Other studies add further support by showing that some neurons in aIT and pIT keep track of the behavioral relevance of visual features independent of their physical properties contributing to control of visual attention [120,121]. Attentional signals found in the temporal cortex were not limited to visual features such as color or form. Many cells in different regions of STS and IT including TEO were responding when attention had to be allocated to the fixation spot [122]. The same neurons were significantly less active when the fixation spot was blanking [122]. These studies provided the earliest evidence that necessary elements for the deployment of visual attention were likely present in IT cortex.

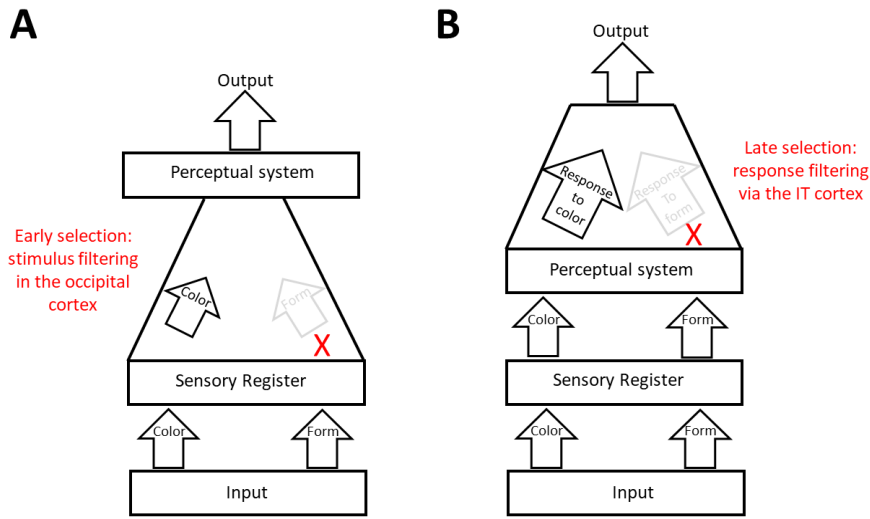

*Figure 1.* Early (A) versus late (B) attentional selection. While both types of filtering reduce the effect of the distractor, early selection mechanisms have a shorter time course and mainly act on the early representation of the stimulus. In contrast, later selection mechanisms filter out the behavioral responses to the distractors, have a longer time course of action and involve higher-order areas.

Neuroimaging studies have also highlighted a potential role for temporal cortex in controlling spatial and feature-based attention in both monkeys and humans [29,123,124]. As discussed in [125], one reason that prior neurophysiological studies may have missed the contribution of the temporal cortex to attentional control is that they mainly focused on single neuron responses rather than how attention is reflected in the population responses. In an attempt to study attentional modulations using population coding, Sereno and Lehky utilized an experimental paradigm in which monkeys had to either pay attention to the location or shape of a stimulus [126]. While single neurons in ventral and dorsal stream areas (aIT and LIP) were significantly modulated by attention, multidimensional scaling analysis at the population level showed a significant attentional effect (better discriminability between locations and shapes) only in the aIT. Hence, the strength of attentional modulation of shape and location at a single-cell level in the temporal cortex might have been underestimated in the past (see [125]). Nevertheless, it is not *only* the strength of modulation in IT cortex that is relevant, but whether attentional control signals originate from it as well.

Evidence for the latter comes from studies showing that ablating of cortical structures that include TEO impairs selective attention [127,128]. Lesioning areas V4 or TEO had a significant effect on the monkeys' ability to filter out distractor information which interfered with the discrimination of targets within several feature domains. These attentional deficits were around two times larger when both V4 and TEO were simultaneously lesioned. The same group later showed that V4 and TEO are essential for successful spatial generalization tasks [127,128]. Furthermore, neural responses in more anterior regions of IT cortex are significantly altered after V4/TEO lesions [129,130]. So, in order to efficiently filter out a distractor which is simultaneously presented together with a target in the receptive of IT neurons requires attentional filtering occurring within earlier areas with smaller receptive fields, in this case, V4 and TEO. As per the Selective Tuning theory of attention, this filtering would incorporate both local circuitry and feedback from later stages of the ventral visual stream [37]. We propose that this feedback originates in higher order areas and backpropagates to earlier visual areas.

**Temporal cortex attention network (TAN)**

This section reviews recent evidence from behavioral, neurophysiological and fMRI studies that revealed distinct regions in the IT cortex and the STS, collectively referred to as the temporal cortex attention network (TAN), that guide visual attention in the presence of cognitive control signals.

*Area pITd*

In a recent monkey fMRI study, area pITd, an area in the dorsal part of the posterior IT cortex, showed strong modulation during an attentive motion processing task [45]. This surprising finding contrasted with the general wisdom that attentional modulation of a ventral stream area is a consequence of drawing attention to the feature that area primarily processes [108,123,131], because area pITd had been shown to be sensitive to shape [95] and color [132], not motion. Despite the fact that motion was the main feature to be attended, motion-sensitive areas were much less modulated in their experiment. In a follow-up electrophysiological experiment [46] they showed that area pITd exhibits the properties of a priority map encoding the spotlight of attention similar to areas LIP and FEF [26,133]. They determined that: (1) three different tasks (motion detection, motion discrimination, and color discrimination) yielded similar attentional patterns. (2) Neurons within the pITd were not tuned to task-relevant stimulus features, like motion direction or color. (3) The activity of the pITd neurons was highly correlated with upcoming errors of attentional selection. (4) Microstimulation of pITd deployed spatial attention. This frames pITd as the first non-oculomotor area that exhibits the properties of a priority map, as described by Fecteau and Munoz [48]. Importantly, area pITd is a *feature-blind* priority map in the ventral visual stream that can guide spatial attention. It is plausible that area pITd combines input from lower order feature conspicuity maps (such as area MT) and higher order areas involved in generation of top-down relevance signals to guide behavior.

A key factor of an attentional priority map is the ability to focus attention to specific locations within as well as across hemifields. As the recent pITd study [46] only contained one target per hemifield, further studies need to show that the area pITd can deploy spatial attention to specific locations within the same hemifield (see [28]) to fully establish pITd as a new priority map.

*The STS area*

Recent studies have delineated two distinct areas in the STS which are implicated in the control of visual attention: (1) an area in the posterior superior temporal sulcus, has been designated as the gaze following patch (GFP) as it has been demonstrated to have a key role in controlling spatial attention during social interactions in humans and monkeys [44,134–

136] and (2) the middle STS region (mid-STS) in monkeys which has been shown to control spatial attention during motion direction and orientation discrimination tasks [42,43].

### 1- GFP

The GFP located in the posterior superior temporal sulcus has been shown to specifically support the faculty of joint attention [134,135,137], a necessary component of primates' social development [138–140]. Ramezanpour and Thier performed the first electrophysiological recordings from the monkey GFP showing that neurons in this area are spatially tuned to locations that someone else is looking at, enabling the observer to shift their attention to the same [44,136]. Furthermore, the study implicated GFP activity in cognitive gaze following behavior, highlighting the role of social context in shaping pSTS neural activity. In a control condition, in which the same visual stimuli were presented to the monkeys but they had to ignore the gaze cues and instead use the portrait's identity information to find the correct target, the same neurons that had been found to be spatially selective for gazed-at locations lost their spatial tuning. This result highlights the role of the GFP in cognitive control of gaze following according to task demands [44,136]. Furthermore, the GFP neurons could target more than one gaze location within each hemifield, a finding which has yet to be shown in area pITd.

These findings not only suggested that the GFP plays a key role in a circuit controlling spatial attention, but they could also show the importance of this area in contextual control of attention. In their experimental paradigm, there was a period in which the animal had to pay attention to the color of a small dot. Different colors implied different cognitive strategies the animal had to take to solve the task, either following the subsequent face's gaze or to map his identity onto the same spatial targets layout which differed with the gazed-at location in most of the trials. The population of neural responses in the GFP was distinctively sensitive to the two instruction colors, red and green, and highlighted the importance of this area in the control of attention based on non-spatial contextual information. Notably, the differential responses to the color of the central fixation were tightly linked to behavioral performance as the animals made significantly more errors when the two separate populations of instruction selective neurons lost their selectivity. These findings show that the GFPis involved in determining the relevance of the social cues such as faces and hands to behavior in nonverbal communications [97] or predictability of social interactions [141].

### 2- mid-STS

Further evidence for the involvement of non-oculomotor structures in the control of spatial attention comes from two recent studies investigating consequences of midbrain SC inactivation on the rest of the brain activity during two covert attention tasks [42,142]. Previous studies of the same group showed that responses in monkeys' middle parts of the STS (mid-STS) and not motion-sensitive areas in the dorsal stream as most strongly attenuated after SC inactivation in an attentive motion discrimination task [43]. This reduction in the attentional modulation was replicated using another task replacing the random dot motion stimuli with second-order orientation stimuli that could not be discriminated according to the changes in motion energy. Finally, they inactivated the mid-STS region directly and observed attentional performance was similarly disrupted. Interestingly, in a follow-up study, the same group showed that many neurons in this area exhibit object selectivity and SC inactivation reduces this selectivity [142]. Altogether, these studies suggest that the mid-STS is involved in the SC's control of spatial attention, and thus is part of the hitherto oculomotor network's control of spatial attention. Interestingly, the mid-STS in monkeys has a functional connectivity fingerprint very similar to the temporoparietal junction (TPJ) in humans [143]. Hence, it's functionality might be reminiscent of the TPJ i.e. reorienting attention towards a novel object of interest [144]. However, the exact correspondence has yet to be confirmed. Nevertheless, the mid-STS attentional modulations are partially necessary for normal object representation in the temporal cortex [142]. It is important to note that the mid-STS area was located a few millimeters anterior to the reported anatomical coordinates of the GFP and pITd areas [44,46]. The question of whether all of these three areas belong to the same cytoarchitectonic structure remains open.

Inactivation of the mid-STS caused monkeys to exhibit "spatial neglect", ignoring stimuli in the contralateral visual field [42]. While the early studies on spatial neglect suggested it was due to lesions to the temporo–parieto–occipital junction and inferior parietal lobule, later studies refined the location to the superior temporal cortex in which lesions cause spatial neglect [145–147]. Thus, the neurophysiological inactivation of mid-STS results in monkeys are consistent with spatial neglect symptoms in humans as a consequence of broader STS lesions.

## What do the TAN regions have in common?

One interesting common denominator of the above mention studies on the TAN is that all of the behavioral tasks have an executive control component i.e. a gaze should be followed or not [44], a lever should be released or not [42], a saccade should be made or not, [46] and all of these functions must be synchronized with temporal requirements of the tasks to ensure flexible performance. When the cognitive control signal interacts with attentional control, the TAN is engaged to support flexible behavior, a finding very similar to what has been already found in area LIP [148,149]. Previously Oristaglio and colleagues had found some neurons in area LIP integrate covert spatial attention *and* a learnt stimulus-action association [148]. In their task, the animals had to release a bar held in their right paw if the cue was oriented to the right or a bar held in the left paw if it was oriented to the left. A considerable number of LIP neurons responded to the attended location, something typical for LIP neurons, and the bar release. As discussed in depth in [149], these neurons might be the basis of target selection interfaces with higher order processes of executive control which facilitate relevance assignment to visual cues. Indeed, the TAN neurons might have inherited their mixed selectivity (spatial attention and executive control) properties from their tandem LIP neurons or vice versa. It should be noted that such integrations are likely dependent on the visual working memory signals, which can contain task-relevant instructions and progression of the representations from the earlier visual areas up to the higher-level areas [150]. Indeed, a study revealed that successfully retaining visual information in working memory depends on a corticocortical loop of the prefrontal and IT cortex since bilateral cooling of each of these areas induced, in the other region, changes of spontaneous and task-related neuronal responses which were accompanied by lower performance in a working memory task [114]. This bidirectional interaction between the temporal cortex and the prefrontal fits well the selective tuning model of attention in which bottom-up signals from lower-level visual areas **and** top-down signals from higher-level areas are both needed to select a target. Another common denominator of the above-alluded studies is that they need to deal with spatial transformations to generate a saccade [151]. By a closer look at the behavioral tasks used in the studies on the TAN, it becomes clear that output spatial coordinates derived from visual processing at the focus of attention (often in the periphery) need to be integrated with extra-retinal signals, such as the current eye position, to enable generating a precise saccadic eye movement initiating from the fixation point. Such spatial transformations were first observed at the level of the parietal cortex via gain modulation [152,153]. This concept, called "gain fields", was later proposed to be important for invariant object recognition with the modulatory quantity being attention and led to the idea that IT cortex may use an attention-centered rather than eye-centered mechanism for invariance in object recognition [154–156]. We predict that the TAN might contribute to such gain field transformations in a more general way as they can deal with spatial processing in parallel to complex features analysis as opposed to parietal neurons which are not involved in object recognition. The importance of spatial transformations, especially in social interactions, has been discussed in depth by Chong (2013) [157].

As noted by Tsotsos and colleagues [64], executive control of visual attention must deal with fixations (overtly or covertly), integration of bottom-up with top-down information (such as rules or prediction signals), spatial localization, priming, and other ingredients of visual tasks while enabling precise timing of the overall behavior. "Cognitive programs", or executive controllers, have been proposed to provide mechanistic integration of the above algorithms to ensure flexible behavior. The tasks recruited in studies of temporal cortex control of attention [42–46,135,142] have several attentional elements such as cueing, priming, selection, covert or overt attention, endogenous or exogenous initiation, spatial localization, disengaging and shifting attention, and surround suppression (for a full list of attentional elements see [63,64]). Hence it is necessary that a cognitive program oversees the whole process. The cognitive programs concept, proposed by Tsotsos and Kruijne [63], is an advanced version of Ullman's visual routines [158] and emphasizes the fact that attention is much more complex than just selecting a region of interest for gaze fixation. Hence, active vision requires a set of algorithms beyond simple relationships between extracting shapes and spatial relationships to ensure it reaches its goal. The recently discovered TAN might indeed be representative cognitive programs for these types of tasks or at least contribute substantially though which exact aspects of cognitive programs are embedded in those areas must be determined through future studies. It is particularly important to investigate whether they are innate or they are developed via learning, where they are stored, and how they are retrieved.

Altogether these findings support the notion that the neural circuits in the TAN are not just passive receivers of attention signals, but they also significantly contribute to executive control of visual attention and implementation of cognitive programs during complex behaviors.

Previously, alternative models for visual attention control in humans, such as dorsal and ventral attention networks (DAN and VAN, [3]) attempted to explain where signals associated with top-down cognitive control might be integrated with bottom-up feature salience. Corbetta and Shulman suggested that while top-down signals for control of visual attention are generated in the dorsal posterior parietal and frontal regions (DAN), the VAN, including temporoparietal junction (TPJ) and inferior frontal cortex, directs attention to salient events in a more bottom-up manner and normal vision requires both systems to interact [3]. A follow-up fMRI study did not find functional evidence of a TPJ in macaque monkeys when testing them using the same paradigms which activated the human TPJ [131]. Interestingly, another study revealed that the mid-STS area in macaques, part of the TAN , has the same functional connectivity profile as the human TPJ [143]. These observations may raise the question of whether the monkey TAN might be indeed homologous to the human VAN. We think this correspondence does not fit the functional properties of the monkey TAN and the human VAN for two main reasons: (1) The human VAN has been shown to be driven mainly by low-level features of the stimulus, such as color, in a bottom-up manner and regardless of the ongoing task, e.g., when they appear outside the cued focus of spatial attention unexpectedly [3]. However, the monkey TAN has been shown to be driven by more complex features, such as other's gaze [44], and cognitive processes at the focus of the attention in a task-dependent manner, such as paying attention covertly to the movement direction of stimulus while ignoring another stimulus in the opposite visual field over a long time [42,46]. (2) Neuroimaging studies have more consistently reported right-hemispheric activation of the VAN (see also for a review [159]), however, the monkey TAN is bilaterally activated.

Of course, the next logical step to investigate if there might be a human TAN, neither overlapping functionally nor anatomically with the human VAN, would be to conduct comparative fMRI studies in humans, now using the same behavioral paradigms hired in monkey studies on the TAN. With the same logic, two studies attempted to localize the TAN regions in the human cortex [135,160]. These two studies revealed that the areas corresponding most closely to the monkey GFP and pITd are clearly far from the human TPJ. The human homologue of the monkey GFP and pITd, constituting the human TAN, are located much more inferior to the human TPJ. Altogether, these observations confirm that there exists a TAN in both monkeys and humans, functionally and anatomically segregated from the posterior member of the VAN, i.e. the TPJ.

While the functional segregations of attention networks (DAN, VAN, and TAN) could explain some of the empirical observations, the mechanism by which these putative networks should interact is not well understood. The "Coherence field" notion has been able to provide a plausible solution [161]. This concept proposes that selective attention synchronizes the activity of neurons across topographically organized stages of the hierarchy to form coherence fields in which different areas contribute complementary information to support target selection. The coherence fields also support the notion that the relative influence of bottom-up stimulus features and top-down behavioral goals on the concerted activity of neurons in the visual hierarchy varies along a continuum. Therefore, findings an exact locus in the hierarchy at which a shift from "source" to "target" occurs is complex [161]. This idea fits the selective tuning model (Tsotsos et al., 1995), which implies that the focus of attention is present throughout the whole neural network, starting at V1 and ending at the frontal cortex but with different resolutions (spatial and feature-based) and recurrent processing.

## What could explain the functional properties of the TAN?

### 1- Feature specialized circuits

The anatomical location of the strong attentional signals found in the previously described recent studies of the mid-STS, GFP, and pITd is not identical, albeit very close [42,44,46]. What mechanisms could explain the strong attentional control signals found in these studies and not in more anterior parts of the temporal cortex such as aIT?

The first important element is the degree of these areas' (or their immediate neighbors) specialization with respect to different behavioral tasks employed. Emergence of highly specialized circuits to optimize neural information processing has been shown to be tightly linked to brain topology [162] and the temporal cortex is not an exception. Current hypotheses suggest that at least the GFP is a domain-specific module since it controls spatial attention only based on social gaze information (see [44] for discussions). Whereas the pITd or mid-STS areas have been shown to be more generic as they respond to a wider range of features (such as motion and color). While the full characteristics of each area specialization needs to be explored in more detail, importantly they are all adjacent to areas specialized with respect to the feature of the task being performed. For example GFP, with its dependence on social gaze, is in between posterior and middle face patches [134]. The same holds for areas pITd and mid-STS which are sitting next to motion (MT, MST and FST) processing areas [43,45]. This close proximity to feature conspicuity maps may empower the attention control areas to access and integrate the necessary streams of information via local circuitries. Similarly, one can speculate that paying attention to highly complex shapes such as a face identity may recruit more anterior regions of the IT cortex specialized in facial identity processing. As a matter of fact, an fMRI study has already shown that when monkeys were instructed to use other monkeys' facial identities to pinpoint particular spatial targets associated with those identities, an area in the anterior temporal cortex, likely one of the anterior face processing patches, was activated [134].

Still two questions remain for future studies: (1) what is the relationship between feature selectivity in the TAN regions and their attentional control strength? (2) Are the attention signals observed in the temporal cortex driving spatial attention (location) or object-based attention (the object itself)? The first question can only be answered by conducting experiments using a battery of visual tasks across a range of features. The second question can be addressed by dissociating spatial and object-based attention in a scenario such as priming spatial attention in the absence of visual information (see [163] as an example).

We should emphasize that while we presented the findings of the recent studies as three separate areas, there is converging evidence that specific cognitive operations are emergent property of network operations rather than being strictly linked to activity in restricted parts of the brain. Future studies are required to investigate how these areas interact with the rest of the brain, including the VAN and DAN, during cognitive operations.

## 2- Visual field maps

A visual field map is a representation that might be used for control mechanisms and could explain the distinct loci of attentional control signals beyond different behavioral tasks (and different features). Early imaging studies on the retinotopic organization of the visual cortex did not consistently find retinotopy in IT cortex. The initial lack of retinotopic maps in other parts of the brain other than early visual areas could have been either due to the low signal to noise ratio of the neuroimaging technology or not using an appropriate protocol which considers stimulus-based selectivity in addition to the visual field location [164]. However, this notion was later refined by some studies showing a systematic representation of eccentricity in the ventral stream [84,165–169]. To date, the existence of retinotopic maps have been shown in parietal, and even frontal cortex [170,171]. These findings are important as one can hypothesize that retinotopy might constitute a basis for connecting different parts of the frontal, parietal and temporal cortex dealing with the same part of the visual field together.

Note that in each of the above studies attention had to be deployed to a certain location in the visual field. In the studies on the pITd the two stimuli were presented at five degree eccentricity [45,46], while in the studies on the mid-STS the stimuli were presented at eight degree eccentricity [42,43], and the GFP study included a central stimulus which had to be attended foveally [44]. Apart from other behavioral demands which varied across these studies, the eccentricity perhaps had played a major role in the location of attention-related activities in the temporal cortex. The fact that the GFP area was very close to the posterior part of the temporal cortex which has a foveal bias could be because they presented the face stimuli at the center of the visual field (see **Figure 2**). Similarly, the pITd and mid-STS studies used more eccentric stimuli, which could have caused a shift in the locus of the attentional modulation according to the visual field representation and eccentricity gradient in the posterior temporal cortex towards more anterior and ventral parts of the temporal cortex (see [73] for a review). If the stimulus eccentricity drive the cortical activation pattern, then one can predict that using more eccentric locations for stimulus presentation may shift the locus of attentional control even further anterior in the temporal cortex. Future experiments are needed to test this hypothesis.

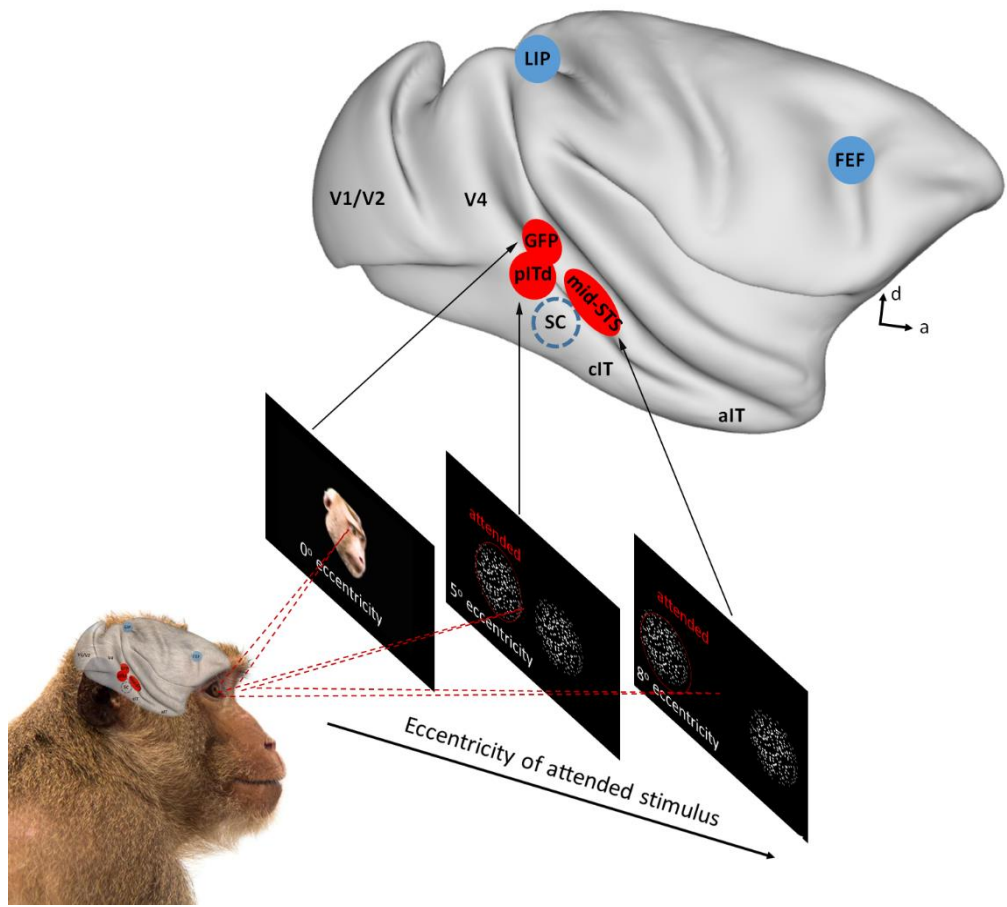

*Figure 2.* Oculomotor and temporal cortex areas involved in the control of attention. Blue: The oculomotor system priority maps, red: The temporal cortex attention modules [42,44,46]. The presentation of the visual stimuli that required to be attended at various eccentric locations might have shifted the locus of attention control signals in the posterior-middle temporal cortex. Note that the locus of priority maps are approximate locations based on coordinates found in the corresponding original papers. To confirm the exact relationship between attentional foci in the temporal cortex and the type and eccentricity of stimuli, future studies should carry out mapping these areas in the same animals.

As described in the selective tuning model, spatial attention is distributed both within a given layer (area within the visual processing hierarchy) as well as feeding back from higher level areas. The existence of retinotopic maps in the frontal and parietal cortices [170] might suggest that areas with the same eccentricity bias in prefrontal and parietal cortices may drive visuospatial attention in the posterior-middle temporal areas which in turn modulate the lower level visual areas even as early as V1 in a top-down manner. In this framework, the temporal cortex attentional signals can be considered as intermediate feedback signals necessary for the formation of high spatial resolution focus of attention in earlier retinotopic areas of the visual hierarchy. In fact, the distinct topography of the attentional control areas seen in the posterior-middle temporal cortex could also be a consequence of the annulus surround suppression as described in the selective tuning model [37]. Each stimulus is presented and attended at a certain eccentricity which suppresses the immediate part of visual field around it in a retinotopic manner. Hence the neighboring areas in the retinotopic map are suppressed (**Figure 3**). That would explain GFP suppression of the neighboring areas pITd and mid-STS in fMRI activation maps. However, future studies are needed to perform all of these tasks and retinotopic mapping on the same set of subjects to confirm whether the GFP, pITd, and mid-STS areas together form a complete map of the visual field.

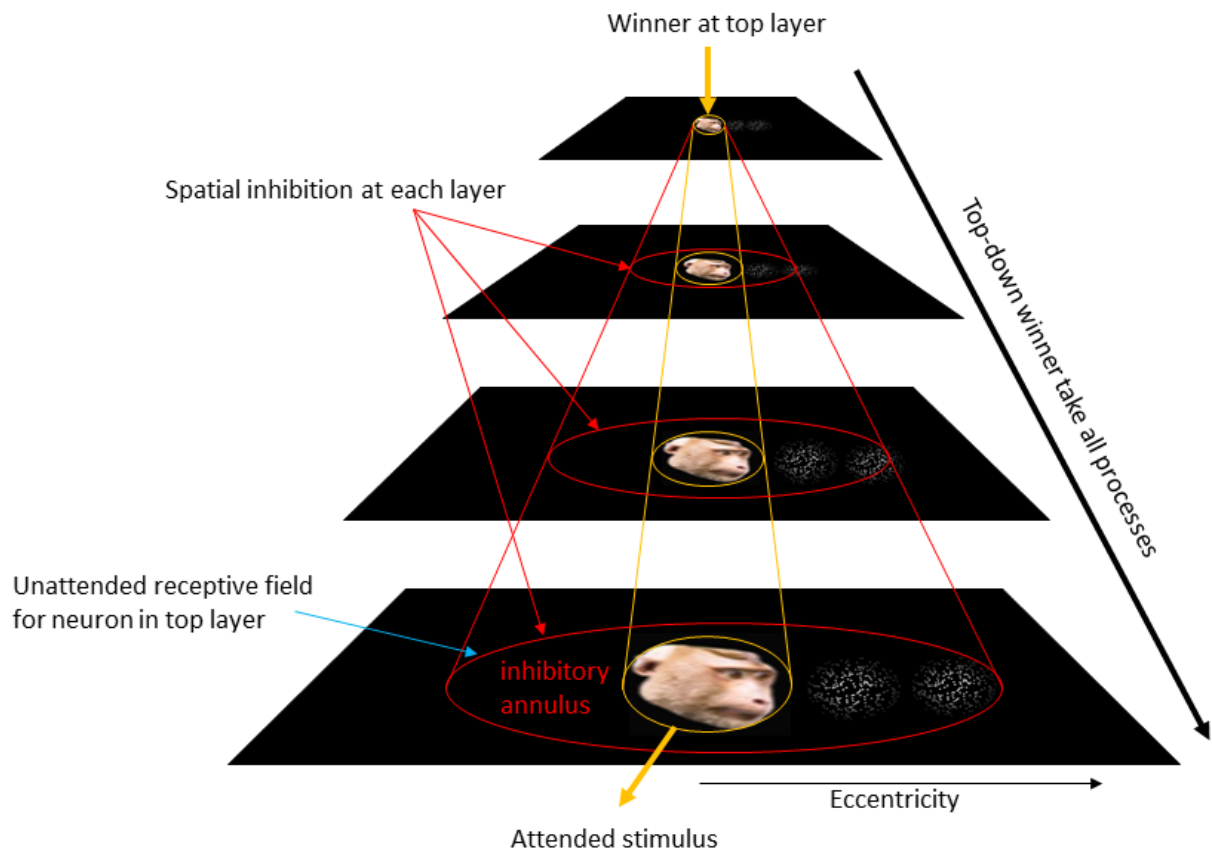

**Figure 3.** Attentional surround suppression. Attending to a face presented foveally activates the GFP and generates a suppressive surround for stimuli presented at larger eccentricities and deactivates the neighboring pITd and mid-STS areas.

## Connectivity of the monkey TAN

The cortical and subcortical connectome of the TAN, specifically to the two other classical attention networks (DAN and VAN) has yet to be fully established. Nevertheless, there is already some evidence which might help to better understanding the TAN's wiring to the rest of the brain. In monkeys, pITd's connectivity further support its ability to be part of the network controlling spatial attention. Diffusion tensor imaging (DTI) showed that while matter bundles originating in pITd are well connected to the two nodes of the DAN: FEF and LIP [172]. Tract tracing studies have shown that the anatomically defined area pITd has feedback projections to earlier visual areas such as V4, V3, V2 and even V1 via either V4-V3 or V4-V2 [173]. Since the GFP is very close to the pITd, one might expect that their connections should not be drastically different than the pITd. In addition to reciprocal connections to the early visual areas, the GFP receives input from a subcortical pathway including the pulvinar [174] and superior colliculus [42,142] and in turn projects to the interparietal sulcus, likely area LIP [175]. While anatomical studies using tract-tracing are informative, they largely depend on the coarse parcellation of cytoarchitectonic areas. Moreover, it is often difficult to conclude whether an anatomically defined area may exactly fit to its fMRI-delineated functional correspondence. One method which can overcome these limitations is combined electrical stimulation-fMRI [176,177]. While concurrent electrical stimulation-fMRI of the functionally defined TAN has yet not been performed, a recent study focusing on the lateral prefrontal cortex (LPFC) revealed that there exists a topographic progression of the LPFC connections on the cortical surface along a particular direction [177] such that a caudoventral to rostroventral gradient in temporal cortex and a caudal/sulcal to rostral/superficial gradient in posterior parietal cortex seem to map to a largely overlapping topographical map in the LPFC. Based on this map the GFP and the mid-STS are likely connected to FEF and its anterior/ventral neighbor. The non-retinotopic continuum in the LPFC which integrates retinotopic maps of the temporal and parietal cortices, can fulfill the requirements of the top layer in the selective tuning model to establish the flow of top-down signals for guiding attention. The TAN as an intermediate level of the attentional control hierarchy, can continuously link early representations in the visual cortex and subcortical areas which are more robust with higher level flexible cognitive processes in the parietal and prefrontal cortices to ensure an optimal behavior.

## Implications of active vision in temporal cortex for social interactions

Intelligent social behavior requires flexibly attending to cues from multiple sensory modalities provided by the other individual. For instance, in one moment we may assess the other's gaze direction in order to identify his/her focus of attention, whereas, in the next moment we may focus on his/her hand pointing toward a certain object being discussed. There is converging evidence that the primate brain treats social information differently to the extent that the existence of a third visual pathway has been recently hypothesized [178]. This third visual pathway which connects the early visual areas to the STS, plays a crucial role in processing dynamic aspects of social perception such as moving faces and bodies, ultimately leading to the understanding of others' actions and theory of mind [179].

We hypothesize that participation of the TAN regions in attentional control and cognitive programs are also beneficial for action understanding. In order to arrive at a complete interpretation of the given social context, we need to flexibly switch between various cues and actions and concatenate the information collected. Hitherto, the neurophysiological principles that orchestrate ensembles of neurons to flexibly link these cues to generate a meaningful and dynamic percept of the other's actions have been mostly studied in the context of mirror neurons and premotor cortex [180]. Nevertheless, there are anecdotal reports of the neurons in the STS representing others' actions [181] similar to what has previously been found in the premotor cortex. It is parsimonious that the areas selective for social cues such as eyes, faces, and hands would also interpret these stimuli as social cues and allocate attention accordingly.

How can the TAN contribute to action understanding during social interactions? Primates as social species need to process and direct attention based on social cues as much as nonsocial ones (such as a flashing red light). At a lower level, attention has been shown to play an important role in binding visual features such as color and motion into an object representation [47,182]. Hence it is parsimonious to assume that attention might also be necessary to bind different social cues provided by others in order to generate a unified action. Take gaze following as an example in which attention should be constantly paid to the other person face in order to be able to detect abrupt changes in his/her gaze direction, head movements, and finally the object he/she fixates on. As described previously, the processes which are needed to perform these actions rely on cognitive programs which the temporal cortex contributes to. Consistent with the selective tuning model of attention, temporal cortex, as an intermediate level in the visual hierarchy, can reduce interference among other elements by operations such as pruning away task irrelevant information.

Why is the temporal cortex a good candidate for implementing the above operations? The essential elements of social interactions such as faces, body parts, and biological motion are indeed already represented in the temporal cortex. Hence implementing attention control signals and cognitive programs at the level of temporal cortex is ecologically beneficial because the local information processing mechanisms already available in these areas (such as lateral inhibition, surround suppression, etc.) can be recruited for binding social cues, flexibly switching between them, and implementing them into attentional control.

**Conclusions**

Through reviewing the latest findings on the potential role of temporal cortex in guiding visual attention, we propose that the STS and IT cortex participate in target selection *and* cognitive programs. We reviewed how different behavioral tasks with different visual stimuli presented at different eccentricities may explain attentional control signals observed in the posterior and middle STS and IT cortex regions, collectively referred to them as the TAN.

However, as this perspective on active vision by temporal cortex generalizes the existence of attention controllers to non-oculomotor structures, new questions arise that cannot be addressed without further study. Perhaps the most important question is what differences, if any, exist between the temporal cortex control of attention and previously known oculomotor system attention control areas. Most of the previous studies which missed attentional control signals in the temporal cortex had used relatively simple tasks and focused on control areas that produce motor output since they are easier to compare to behavior. However, it is becoming more evident that complex object representations can guide motor systems which do not represent that complexity (see [183] as an example from the oculomotor system). This notion supports the existence of cognitive programs in temporal cortex which are necessary to work in conjunction with oculomotor priority maps in frontal and parietal cortices to produce complex behaviors. The use of cognitive programs is a parsimonious solution to the brain as a dynamical system which needs to deal with a wide range of complex tasks and behaviors. Different nodes of this system play different roles for specific tasks, while the connectivity ensures the relevant control signals reach their target(s). The area that is most able to differentiate the visual scene into different priorities would be the one to drive the overall system. This view also fits to the selective tuning model of attention [37]: the focus of attention is present throughout the whole neural network starting at V1 and ending at frontal cortex but with different resolutions (spatial and feature-based) and recurrent processing. Control of visual attention is not just a function of the fronto-parietal motor control networks. Temporal cortex also constitutes cognitive programs that deploy visual attention to earlier stages of the ventral stream to support target selection and enable flexible complex behaviors.

**No data statement:** This paper does not contain original data or analysis.

**Declaration of Competing Interest:** None declared.

**Acknowledgment:** H.R. was supported by a Vision: Science to Applications (VISTA) postdoctoral fellowship award. M.F. was supported by an NSERC Discovery Grant and CIHR Project Grant.